\documentclass[a4paper,11pt]{article}

\usepackage{jheppub} 

\usepackage[T1]{fontenc} 

\usepackage{amsmath, amssymb, bm, graphicx, graphics, color, mathrsfs}

\newcommand{\be}{\begin{equation}}
\newcommand{\ee}{\end{equation}}
\newcommand{\ben}{\begin{eqnarray}}
\newcommand{\een}{\end{eqnarray}}

\newcommand{\la}{{\lambda}}

\newcommand{\cO}{{\cal O}}

\newcommand{\cR}{{\cal R}}

\newcommand{\p}{\partial}
\newcommand{\na}{\nabla}

\newcommand{\tr}{{\tilde \rho}}

\newcommand{\ep}{\epsilon}

\title{\boldmath 
P-wave holographic superconductor/insulator phase transitions affected by dark matter sector}

\author[1]{Marek Rogatko\note{rogat@kft.umcs.lublin.pl, marek.rogatko@poczta.umcs.lublin.pl}}
\author[2]{Karol I. Wysokinski\note{karol@tytan.umcs.lublin.pl}}
\affiliation{Institute of Physics \\
Maria Curie-Sk{\l}odowska University \\
20-031 Lublin, pl. Marii Curie-Sk{\l}odowskiej 1, Poland}

\abstract{
The holographic approach to building the p-wave superconductors results in three different
models: the Maxwell-vector, the SU(2) Yang-Mills and the helical. 
In the probe limit approximation, we analytically examine the properties of the first two models
in the theory with {\it dark matter} sector. It turns out that the effect of {\it dark matter} on the Maxwell-vector
p-wave model is the same as on the s-wave superconductor studied earlier. For the non-Abelian 
model we study the phase transitions between p-wave holographic insulator/superconductor and metal/superconductor.
 Studies of marginally stable modes in the theory under consideration
allow us to determine features of p-wave holographic droplet in a constant magnetic field.
The dependence of the superconducting transition temperature on  the coupling 
constant $\alpha$ to the {\it dark matter} sector is affected by the {\it dark matter} density $\rho_D$.
For $\rho_D>\rho$ the transition temperature is a decreasing function of $\alpha$. 
The critical chemical potential $\mu_c$ for the quantum phase transition
between insulator and metal depends on the chemical potential of dark matter $\mu_D$ and for
$\mu_D=0$ is a decreasing function of $\alpha$. }


\begin{document} 
\maketitle
\flushbottom

\section{Introduction}
\label{sec:intro}

The gauge/gravity duality provides a powerful theoretical method which enables a better understanding of 
the strongly coupled systems \cite{mal}-\cite{gub98}.  Originally proposed as
the equivalence between type IIB superstring theory on $AdS_5 \times S^5$ spacetime and ${\cal N}=4~ SU(N)$
supersymmetric Yang Mills theory on $(3+1)$-dimensional boundary, has later been generalized to  other
gravitational backgrounds \cite{rev1}. The correspondence empowers an equality between the quantum field theory
in $d$ -dimensional spacetime and the gravity theory in $(d+1)$-dimensions, and its usefulness
originates in a strong-weak duality \cite{sachdev2012}. Namely, the gravity dual of the strongly coupled quantum 
field theory  is tractable in a perturbative  approach.                  

The AdS/CFT correspondence has been recently proposed as a method to describe superconducting 
phase transition of the single s-wave superconductor \cite{har08}. Shortly afterwords it  has been 
generalized to take into account other symmetries, like simple  p or d wave. 
The generalizations require proper choice of the condensing field and appropriate 
gravity background.  To describe d-wave holographic superconductor the charged massive spin-two 
field in the bulk \cite{che10}-\cite{zen10} is required. The works related to building the holographic p-wave
superconductor \cite{gub08} have indicated a number of equally feasible possibilities and resulted in
the multitude of approaches to study them. {\it Inter alia} the five-dimensional supergravity framework \cite{apr11}, and  
the Sturm-Liouiville eigenvalue problem \cite{gan12} have been applied. A handful of  novel results  \cite{bas10,amm10,liu15}
have also been reported. 
Especially intriguing result \cite{amm10} is the change of the order of the superconducting transition 
showing up when backreaction is taken into account.  The second order phase transition is
replaced  by the first order one, when matter field couplings are beyond a critical value.

Recently, the aforementioned studies were generalized in many other ways. The modification of gravity theory by considering
the five-dimensional AdS solitonic metric has been proposed \cite{hor98}. It enabled the construction  of the holographic
insulator/superconductor phase transition at zero temperature \cite{nis10}. Namely, the AdS soliton line element
dual to a confined field theory with a mass gap, imitates an insulator phase \cite{wit98a}. 
The strength of various kinds of matter backreactions has been shown to generate new phase 
transitions \cite{hor10,bri11}.

The marginally stable modes of scalar and vector perturbations of the AdS spacetime
have revealed the outset of the phase transition and help to study the influence of magnetic field on them
\cite{cai11a,cai11b}. In agreement with known phenomenology the magnetic field in the holographic theory 
makes  the phase transition harder to occur. Both analytical and numerical methods devoted to the properties 
of s-wave and p-wave insulator/superconductor phase transitions were investigated in \cite{cai11c,akh11}.
The studies in Gauss-Bonnet gravity were presented in \cite{pan11,cai11back},  the effects of the Weyl corrections 
on p-wave holographic phase transitions  \cite{cai13} were studied 
in \cite{zha15}. The p-wave holographic superconductors in different gravity backgrounds \cite{cha15},
and in the presence of non-linear electrodynamics \cite{zha13,jin12}  and other non-trivial conditions 
\cite{alb09,roy13,amo14} have  been elaborated.

The important problem being of  special interest for the present work is a possible matter configuration in the Universe.
According to numerous studies more than 24\% of the matter is invisible and therefore dubbed as {\it dark matter}.
There exist various proposals of how to model this component of matter. In this paper the point of view has been accepted 
according to which the {\it dark matter} is described by the U(1) field \cite{bri11} analogous 
to the Maxwell  one and coupled to the
ordinary matter. The coupling constant $\alpha$ is treated as a free parameter, which value
is bounded by $\mid\alpha\mid<2$.

This model of {\it dark matter} is supported by numerous astrophysical observations \cite{integral,atic,pamela,massey15a,massey15b}
and other experimental data  related to the muon anomalous magnetic moment \cite{muon}
and experimental searching for the 'dark photon' \cite{afa09,gni08,suz15, mir09,red13}.

The main aim of this paper is to study the influence of the {\it dark matter} sector on the properties of
holographic p-wave phase transitions.  These studies may hopefully result in discovery of some 
qualitatively untypical behavior which could be tested in 
the future experiments elucidating the cloven nature of {\it dark matter}.

The superconducting transition  is signaled by spontaneous breaking of the $U(1)$-gauge symmetry.
In the case under consideration the rotational symmetry is also broken by a special direction of a vector field, which 
is obtained by the condensation of a charged vector field. The present paper is the generalization of
 the previous works \cite{nak14,nak15,nak15a},  were   various aspects of phase transitions 
in s-wave holographic superconductor theory  with ordinary matter sector coupled to the {\it dark matter} 
one have been elaborated. 
One expects that {\it dark matter} authorizes a part of a larger particle sector interacting with 
the visible matter  and not completely decoupled \cite{vac91}-\cite{dav13}.

The main point of our studies is the question how the {\it dark matter} sector modifies 
the ordinary phase transitions known from the previous studies of p-wave superconductors. The key role will be played 
by the dark matter coupling constant $\alpha$, binding dark matter fields  with the ordinary Maxwell gauge 
field. It is important to know how the phase transitions are modified by the {\it dark matter} sector.
To this end the holographic p-wave  metal/superconductor and holographic insulator/p-wave 
superconductor phase transitions as well as p-wave holographic droplet embedded in magnetic field have 
been studied. As discussed in detail in section 2 the holographic p-wave superconductors provide a numerous
non-trivial possibilities to look for the influence of the {\it dark matter} on them. Out of three 
possible models of holographic  p-wave superconductors we shall discuss in detail two of them.

The paper is organized as follows. In section 2 we describe two of the three  possible models used
to build the p-wave symmetry superconductors and conclude that the effect of {\it dark matter} in one of them is 
formally the same as for s-wave symmetry. This prompts us to consider the other model for which
this influence is much more interesting. In the subsection B we discuss the crucial points of the considered
p-wave holographic superconductor model with the influence of the {\it dark matter} sector. In section 3,
as a gravity background we assume five-dimensional AdS Schwarzschild line element and 
study the metal/superconductor phase transition. {\it Dark matter} sector effects on the 
insulator/holographic p-wave superconductor are analyzed in section 4, while
section 5 is devoted to p-wave holographic droplet in the presence of a constant magnetic field.
We conclude our researches in section 6, paying attention to the new features of the elaborated
phenomena induced by the presence of {\it dark matter} sector which can potentially serve as an
indicator for the future experiments dedicated to the detection of {\it dark matter}.

\section{Models of p-wave holographic superconductor with {\it dark matter} sector}
\label{sec:models of p-wave}
In the literature on the subject there exist at least three different possible ways 
of building p-wave holographic superconductor. Namely, the Maxwell vector model \cite{cai11b}, 
the SU(2) Yang-Mills \cite{gub08,amm10} one and the helical p-wave model \cite{hel1,hel2}. 
The contemporary  review of the subject can be found in \cite{cai15}.
Contrary to this, the quantum field theory approach (the weak coupling) 
delivers the unique description of p-wave superconductors, under the
condition that one ought only to preserve the required symmetry. 
Namely, from the fact that superconductivity is related to a pairing of the two fermions 
it follows that the total wave function of the Cooper pair has to be antisymmetric, 
with respect to their exchanges. The antisymmetry of the spin part requires
a symmetric orbital part of the wave function and one ends up with s-wave or d-wave superconductors. 
On the other hand, p-wave symmetry of the orbital part of the wave function requires 
the triplet character of the spin part. On the technical level, the field theoretical description of all types of 
superconductors stems from the same type of BCS-like equations which only differ by the symmetry of the
form-factors, $g({\bf k})=1$ for s-wave and $g({\bf k}) =k_x$, where {\bf k} is a wave-vector, for the simplest
p-wave symmetry.
At present, it is not clear which of the 
aforementioned holographic models is the proper one for the description of 
strongly coupled p-wave superconductors and what are the differences between their 
properties. 

In this paper we shall study the Maxwell vector and the SU(2) Yang-Mills models.
Our analysis relies on the theory in which  the gravitational action is given by
\be
S_{g} = \int \sqrt{-g}~ d^5 x~  \bigg( R - 2\Lambda \bigg), 
\ee
where $\Lambda = 6/ L^2$ stands for the cosmological constant, while $L$ is the radius of the considered AdS spacetime.

Before we proceed to the main subject of the paper let us give some remarks about the 
other model of p-wave holographic superconductor, the so-called Maxwell vector model presented
in \cite{cai13}, supplemented by the {\it dark matter} sector. 
The form of the action is in a close resemblance of the quantum electrodynamical $\rho$-meson, without
irrelevant neutral part of it \cite{dju05}.

The gravitational part is the same as presented earlier, while the matter sector will 
be provided by the action
\ben
\label{s_matter1}
S_{m} = \int \sqrt{-g}~ d^5x  \bigg( 
- \frac{1}{4}F_{\mu \nu} F^{\mu \nu } &-& 
\frac{1}{4}B_{\mu \nu} B^{\mu \nu } - \frac{\alpha}{4}~B_{\mu \nu} F^{\mu \nu } \\ \nonumber
- \frac{1}{2}~\rho^{\dagger}_{\mu \nu}~\rho ^{\mu \nu} 
&-& m^2~\rho^{\dagger}_\mu~\rho^{\mu} + i~q~\gamma_0~\rho_\mu~\rho^{\dagger}_\nu~F^{\mu \nu}
\bigg),
\een
where a complex vector field $\rho_\mu$ with mass $m$ and the charge $q$ was introduced.
$\rho_{\mu \nu}$ is defined by the covariant derivative $D_\mu = \na_\mu - iqA_\mu$ in the form given by
\be
\rho_{\mu \nu} = D_\mu \rho_\nu -  D_\nu \rho_\mu.
\ee
The last term in equation (\ref{s_matter1}) describes the magnetic moment of the vector field $\rho_\mu$. 
The vector field constitutes a charged $U(1)$-gauge field and on the AdS/CFT side
is dual to an operator carrying the same charge under the symmetry in question. A vacuum expectation 
value of this operator will be subject to the spontaneous
$U(1)$ symmetry breaking. The condensate of the dual operator
breaks the $U(1)$ symmetry and moreover because of the fact that we have to do with vector fields, 
the rotational symmetry is broken by choosing a special direction.
In the light of the above claims, the vector field is treated as an order parameter 
and the model  mimics p-wave superconductor.

Further, let us assume that we shall elaborate real vector field to have the connection with the results
obtained in our previous studies. 
By the direct calculations it can be checked that, if we assume that the condensate picks out 
$x$-direction and $U(1)$-gauge fields  have only $t$-components
\be
\rho_\alpha~dx^\alpha = \rho_x~dx, \qquad
A_\mu~dx^\mu = \phi(r)~dt, \qquad 
B_\nu~dx^\nu = \eta(r)~dt,
\ee
the underlying Maxwell vector p-wave holographic superconductor with {\it dark matter} 
sector and with the real components of the vector field give us the same 
description of the phase transitions as the s-wave model studied earlier \cite{nak15a}. All the equations 
of motion are of the same forms when we exchange $\psi(r)$ (which acts as an order
parameter in s-wave case) for the $\rho_x$-component of the vector field. Due to this fact
 we shall not elaborate the vector model and concentrate on the SU(2) Yang-Mills one in the following sections.

\subsection{Model of SU(2) Yang-Mills p-wave holographic superconductor with dark matter sector}
In this section we shall describe the basic features of the SU(2) Yang-Mills holographic p-wave
superconductor model with the {\it dark matter} sector. To begin with
one makes ansatz of an $SU(2)$ Yang-Mills field and two $U(1)$ subgroups of the $SU(2)$, considered as
the ordinary Maxwell one and the other supposed to describe the {\it dark matter} sector, coupled to the Maxwell
electrodynamics. Next, a gauge boson generated by the other $SU(2)$ generator and charged under $U(1)$ Maxwell
subgroup will be taken into account.
 On the other hand, the matter sector is provided by the action
\be
\label{s_matter}
S_{m} = \int \sqrt{-g}~ d^5x  \bigg( 
- \frac{1}{4}F_{\mu \nu}{}{}^{(a)} F^{\mu \nu (a)} - 
\frac{1}{4}B_{\mu \nu}{}{}^{(a)} B^{\mu \nu (a)} - \frac{\alpha}{4}~B_{\mu \nu}{}{}^{(a)}F^{\mu \nu (a)} 
\bigg),
\ee
where $F_{\mu \nu}{}{}^{(a)}$ and $B_{\mu \nu}{}{}^{(a)}$ are two $SU(2)$ Yang-Mills field strengths of the form
$F_{\mu \nu}{}{}^{(a)} = \na_\mu A^{(a)}_\nu - \na_\nu A^{(a)}_\mu + \ep^{abc}~A^{(b)}_\mu A^{(c)}_\nu$.
The totally antisymmetric tensor is set as $\ep^{123} =1$. The components of the gauge fields are bounded with
the three generators of the $SU(2)$ algebra by the relations $A =  A^{(a)}_\beta~\tau^a~dx^\beta$, where
$[\tau^a,~\tau^b] = \ep^{abc}~\tau^c$. The parameter $\alpha$ describes the coupling between ordinary and dark matter $U(1)$-gauge fields.

The equations of motion imply
\be
\na_\mu B^{\mu \nu (a)} + \frac{\alpha}{2}\na_\mu  F^{\mu \nu (a)} + \ep^{abc}~B_\mu{}{}^{(b)}~ B^{\mu \nu (c)}
+ \frac{\alpha}{2} \ep^{abc}~B_\mu{}{}^{(b)}~ F^{\mu \nu (c)} = 0.
\label{e1}
\ee
and for $F_{\mu \nu}$ are provided by
\be
\na_\mu F^{\mu \nu (a)} + \frac{\alpha}{2}\na_\mu  B^{\mu \nu (a)} + \ep^{abc}~A_\mu{}{}^{(b)}~ F^{\mu \nu (c)}
+ \frac{\alpha}{2}~\ep^{abc}~A_\mu{}{}^{(b)}~ B^{\mu \nu (c)} = 0.
\label{e2}
\ee
In order to simplify the above equations we multiply relation (\ref{e1}) by $\alpha/2$ and extract the term $\frac{\alpha}{2} \na_\mu B^{\mu \nu (a)}$. The second term in the equation (\ref{e2})
is replaced by the aforementioned outcome. The final result may be written as
\ben \label{e3}
\tilde \alpha ~\na_\mu F^{\mu \nu (a)} &-& 
\frac{\alpha}{2}
\ep^{abc}~B_\mu{}{}^{(b)}~ B^{\mu \nu (c)}
- \frac{\alpha^2}{4}\ep^{abc}~B_\mu{}{}^{(b)}~ F^{\mu \nu (c)} \\ \nonumber
&+& \ep^{abc}~A_\mu{}{}^{(b)}~ F^{\mu \nu (c)} +
\frac{\alpha}{2}
\ep^{abc}~A_\mu{}{}^{(b)}~ B^{\mu \nu (c)} = 0,
\een
where $\tilde{\alpha} = 1 - \frac{\alpha^2}{4}$.

Both $SU(2)$ Yang-Mills fields, $A_\mu{}{}^{(b)}$ and $B_\mu{}{}^{(b)}$, are dual to some current operators in the four-dimensional boundary field theory. In order to accomplish a
p-wave superconductor with {\it dark matter} sector we postulate that the following is satisfied
\ben  \label{ch}
A &=& \phi(r)~\tau^3~dt + w(r)~\tau^1~dx,\\
B &=& \eta(r)~\tau^3~dt.
\een
In the above relations the $U(1)$ subgroups of $SU(2)$ group generated by $\tau^3$ are 
identified with the electromagnetic $U(1)$-gauge field ($\phi(r)$) and the other $U(1)$ group connected with
the {\it dark matter} sector field ($\eta(r)$) coupled to the Maxwell one. The gauge boson field ($w(r)$) 
having the nonzero component  along $x$-direction is charged under $A_t^{(3)} = \phi(r)$.
According to the AdS/CFT dictionary, $\phi(r)$ is dual to the chemical potential on the boundary, 
whereas $w(r)$ is dual  to $x$-component of a charged vector operator.
The condensation of $w(r)$ field will spontaneously break the $U(1)$ symmetry and is subject 
to the superconductor phase transition.  It breaks the rotational symmetry by making the $x$-direction  a special one
and inclines the phase transition. The transition in question is interpreted as a p-wave superconducting 
phase transition on the boundary. As far as the $U(1)$-gauge field bounded with the {\it dark matter} sector 
is concerned, it has the component $B_t^{(3)}$  dual to a current operator on the boundary.

One can remark that the choice described by the relation (\ref{ch}) is the only consistent choice 
of the gauge field components allowing the analytic treatment of the problem. The direct calculations reveal that
the $x(1)$ and $t(3)$ components of the equation (\ref{e1}) are given as follows:
\ben 
\frac{\alpha}{2}~\na_\mu F^{\mu x (1)} &+& \frac{\alpha}{2}~\ep^{132}~B_t{}{}^{(3)}~F^{tx (2)} = 0,\\ \label{eta}
\na_{\mu} B^{\mu t (3)} &+& \frac{\alpha}{2}~\na_\nu F^{\nu t (3)} = 0,
\een
while the same components of the equation (\ref{e2}) imply
\ben
\na_\mu F^{\mu x (1)} &+& \ep^{1 b c}~A_\mu {}{}^{(b)}~F^{\mu x (c)} = 0,\\
\na_\nu F^{\nu t (3)} &+& \frac{\alpha}{2}~\na_\nu B^{\nu t (3)} + \ep^{3 b c}~A_\mu{}{}^{(b)}~F^{\mu t (c)} = 0,
\een
and the main relation  (\ref{e3}) reduces to the following:
\ben
\tilde \alpha ~\na_\mu F^{\mu x (1)} &-& \frac{\alpha^2}{4}~\ep^{132}~B_t^{}{}^{(3)}~F^{tx (2)} + 
\ep^{1 b c} ~A_\mu{}{}^{(b)}~F^{\mu x (c)} = 0,\\
\tilde \alpha ~\na_\mu F^{\mu t (3)} &+& \ep^{3 b c}~A_\mu{}{}^{(b)}~F^{\mu t (c)} = 0.
\een
We take into account the $x(1)$ and $t(3)$-components of the aforementioned equation.

\section{P - wave holographic metal/superconductor phase transition}
\label{sec:metal sup}
In order to analytically investigate metal/superconductor
phase transition we shall implement the Sturm-Liouville method, which for the first
time has been successfully used in holographic phase transition studies~\cite{sio10}.
To commence with, one considers the background of five-dimensional black hole given by the line element
\be
ds^2 = - g(r)~dt^2 + \frac{dr^2}{g(r)} + \frac{r^2}{L^2}~(dx^2 + dy^2 + dz^2),
\ee
where $g(r) = r^2/L^2 - r_+^4/r^2 L^2$. Without loss of generality we set $L=1$.

The Hawking temperature for the black hole in question, is equal to $T_{BH} = r_+/\pi$. 
The $t~(3)$ and $x~(1)$ components of the equation (\ref{e3})
are provided by the following relations:
\ben \label{e4}
\p^2_r \phi(r) &+& \frac{3}{r} ~\p_r \phi(r) - \frac{w^2(r)}{\tilde \alpha~r^2~g}~\phi(r) =0, \\ \nonumber
\p^2_r w(r) &+& \bigg( \frac{1}{r} + \frac{\p_r g}{g} \bigg)~\p_r w(r) 
+ \frac{\phi(r)~w(r)}{\tilde \alpha~g^2}~\bigg(\phi (r) - \frac{\alpha^2}{4}~\eta(r) \bigg)=0.
\een
The above set of the differential equations should be completed by the adequate boundary conditions.
We impose that on the black hole event horizon $\phi(r_+) = 0$ and the condensing field  has a finite norm. The last 
condition requires that $w(r_+)$ should also be finite. Consequently, the boundary conditions on 
the event horizon may be cast in the form 
\ben
\label{bc-phi-hor}
\phi(r) &=& \phi_{hor}^{(1)}~\bigg( 1 - \frac{r_+}{r} \bigg) + \dots,\\
w(r) &=& w_{hor}^{(0)} + w_{hor}^{(2)}~\bigg( 1 - \frac{r_+}{r} \bigg)^2 + \dots.
\label{bc-hor}
\een
On the other hand, on the boundary $r\rightarrow \infty$ of the considered spacetime one has 
\be
\phi(r) \rightarrow \mu - \frac{\rho}{r^2}, \qquad
w(r) \rightarrow w^{(0)} + \frac{w^{(2)}}{r^2},
\label{bc-inf}
\ee
where $\mu$ and $\rho$ are dual to the chemical potential and charge density, respectively.
$w^{(0)}$ and $w^{(2)}$ are dual to the source and expectation value of the boundary vector operator.
The requirement of having a normalizable solution implies that one sets the source $w^{(0)}$
equal to zero.  

In $z = r_+/r$ - coordinates the equations of motion have the forms 
\ben \label{met1}
\phi''(z) &-& \frac{1}{z}~\phi'(z) - \frac{w^2(z)}{\tilde \alpha~z^2~g_0}~\phi(z) =0, \\ \label{met2}
w''(z) &+& \bigg( \frac{1}{z} + \frac{g_0'}{g_0} \bigg)~w'(z) + \frac{r_{+}^2\phi (z)[\phi(z) - \frac{\alpha^2}{4}\eta(z)]}
{\tilde \alpha~z^4~g_0^2}~w(z)=0,
\een
where $'$ means derivative with respect to $z$-coordinate and $g_0=1/z^2-z^2$.

For $T \rightarrow T_c$ the condensate is very small $w(z)\rightarrow 0$ and $w^2(z)\approx 0$. The value 
of the horizon radius for the black hole with temperature $T_c$ is denoted by $r_{+c}$. 
The equation (\ref{met1}) for the $\phi$ field reduces to the relation
\be
\phi^{''} - \frac{\phi^{'}}{z} \simeq 0,
\ee
which has the general solution of the form $\phi(z)=c_1+c_2z^2$. Using the boundary conditions 
for it as described by the equations (\ref{bc-phi-hor}) and (\ref{bc-inf}) we 
find $\phi \simeq \rho r_{+c}^{-2}(1 - z^2)$. We need solution for the  function $w(z)$ near
 the boundary $z \rightarrow 0$ of the considered spacetime. We shall seek it in the form
\be
w(z) \mid_{z \rightarrow 0} = <{\cR}>z^2F(z),
\label{appr-wz}
\ee
where we have assumed $w^{(0)}=0$ and denoted $<\cR>=w^{(2)}$. 

To proceed further, we shall consider the equation (\ref{eta}) in order to find the relation between $\phi(z) $ and $\eta(z)$.
It can be easily solved and one gets 
\be
\eta(z) + \frac{\alpha}{2}\phi(z) = - \frac{c_1}{r_+^4}z^4 + c_2,
\label{rel-eta-fi}
\ee
where $c_1$ and $c_2$ are constants. This equation (for $\alpha=0$) shows that the
asymptotic behavior of the field $\eta(z)$ has  the form analogous to $\phi(z)$. 
This is correct as $\eta(z)$ is the counterpart of the field $\phi(z)$ in
the dark sector. Due to the formal symmetry between both fields we
assume the similar boundary conditions for $\eta(z)$. First, we require $\eta(1)=0$. It leads to the relation
\be
\eta(z)=\frac{c_1}{r_{+c}^4}(1-z^4),
\ee
Comparing this dependence with the behavior of $\phi(z)$ close to the boundary ($z\rightarrow 0$), we 
introduce the {\it dark matter} density $\rho_D=c_1/r_{+}^2$ and approximate the last equation
close to $z=0$, by the following:
\be
\eta(z)=\frac{c_1}{r_{+c}^4}(1-z^4)\approx\frac{\rho_D}{r_{+c}^2}(1-z^2).
\ee
This allows us to rewrite Eq. (\ref{rel-eta-fi}) as
\be
\eta(z)  = \frac{\rho_D}{r_{+c}^2}(1-z^2) -  \frac{\alpha}{2}\phi(z).
\label{rel-2}
\ee 
Introducing the above relation into equation (\ref{met2}) and using representation (\ref{appr-wz})
of $w(z)$ in terms of  $F(z)$, we obtain the equation
\be
(p(z)~F^{'}(z))' - q(z)~F(z) + \Lambda^2~r(z)~F(z) = 0,
\ee
where we have set
\ben
p(z) &=& z^5~g_0,\\
q(z) &=& - z^5~g_0~\bigg(\frac{4}{z^2} + \frac{2}{z}~\frac{g_0'}{g_0}\bigg),\\
r(z) &=& \frac{z~(1-z^2)^2}{g_0},
\een
and  denoted 
\be
\Lambda^2 = \rho^2~\bigg[\tilde{\beta}-\frac{\alpha^2}{4}\frac{\rho_D}{\rho}\bigg]~\frac{1}{\tilde{\alpha}~r^6_{+c}},
\ee 
with $\tilde{\beta} $ given by the following expression:
\be
\tilde{\beta} = 1 + \frac{\alpha^3}{8}.
\ee
The equation is solved by means of the Sturm-Liouville method with the function $F(z)$ 
satisfying the conditions $F(0)=1$ and $F'(0)=0$.
The Sturm-Liouville method enables us to find the minimum value of $\Lambda=\Lambda_{min}$ by
minimizing the functional 
\be
\Lambda^2 = \frac{\int_0^1 dz~[F'(z)^2~p(z) + q(z)~F^2(z)]}{\int_0^1 dz~r(z)~F^2(z)}.
\label{SL-l2}
\ee
The trial function has been chosen in the standard form  $F(z) = 1-a~z^2$.
The dependence of $g(r)$ on $r_+$ has been taken into account by our definition of 
$g_0(z)=z^{-2}-z^2$ in the functions $p(z)$, $q(z)$ and $r(z)$. 

\begin{figure}
\includegraphics[width=0.5\linewidth]{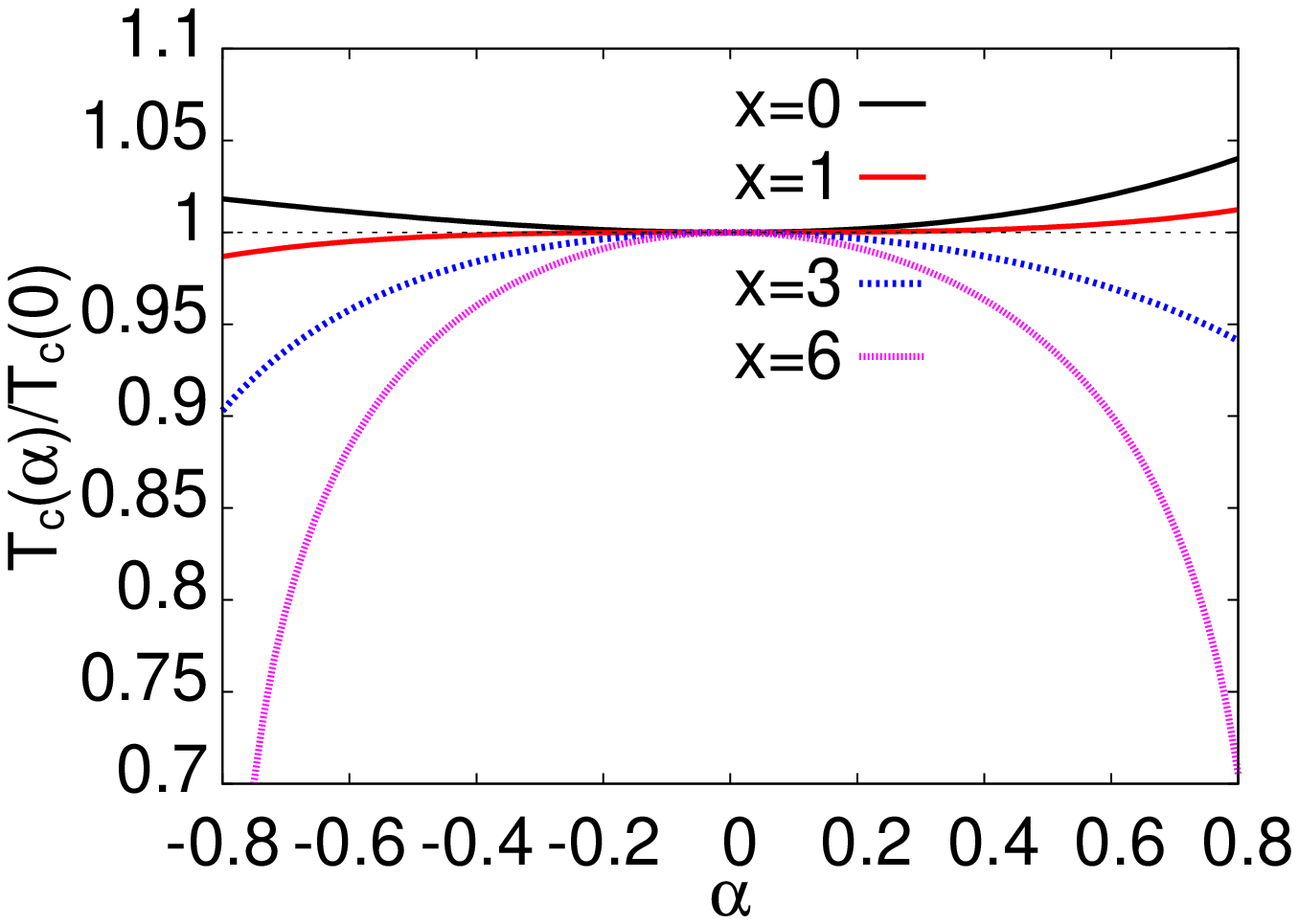}
\includegraphics[width=0.5\linewidth]{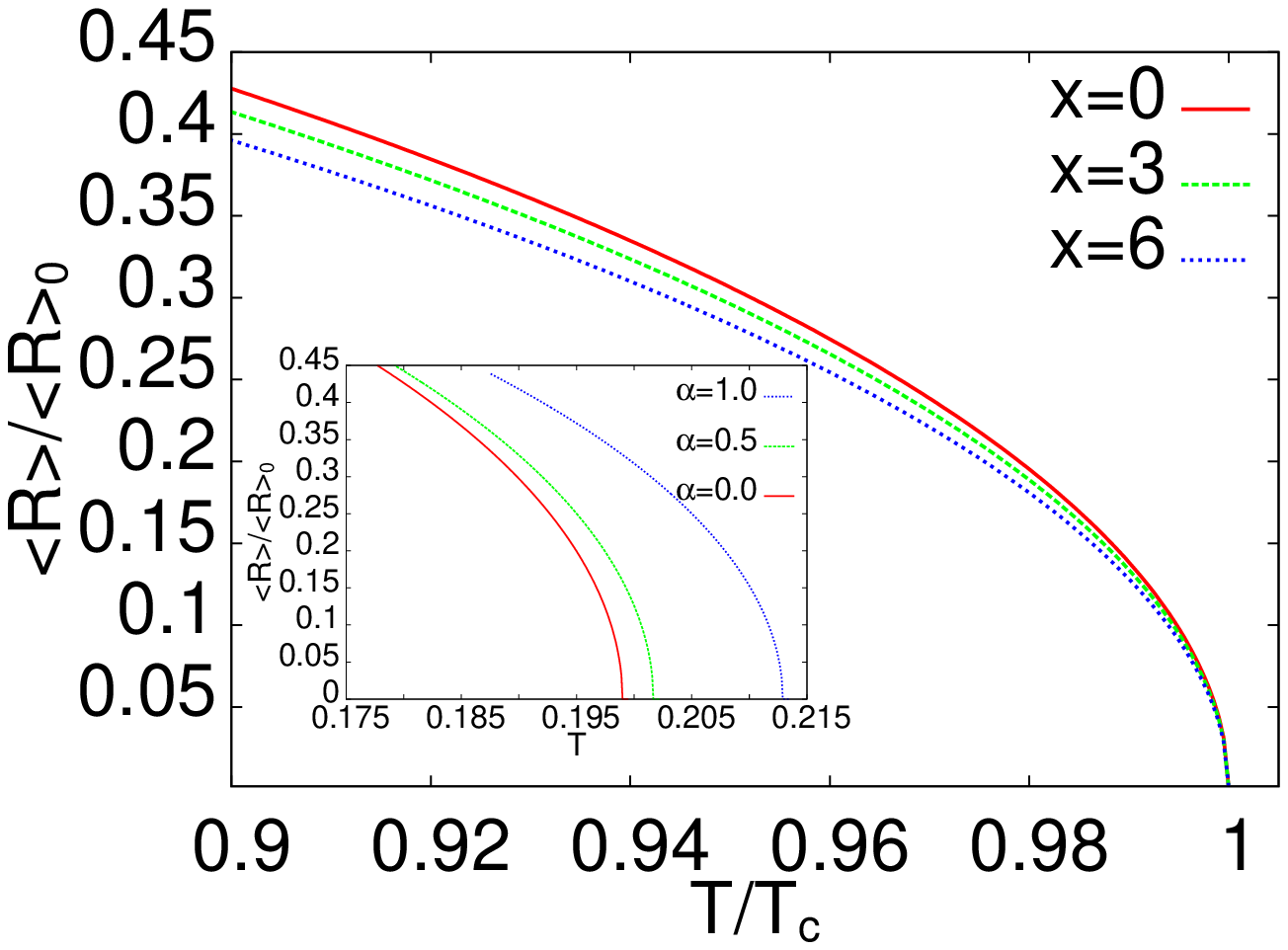}
\caption{(color online)  Left panel shows the  $\alpha$ dependence of the transition 
temperature $T_c(\alpha)/T_c(0)$ normalized to its value without {\it dark matter}.
The right panel depicts the  dependence of the condensation value $<\cR>$ normalized to its
value $<\cR>_0$ calculated for $\alpha=0$ on temperature $T$ normalized to the transition temperature 
$T_c$, for a few values of the ratio $x=\rho_D/\rho$ and $\alpha=0.5$. 
The inset shows the dependence  $<\cR>/<\cR>_0$ on $T$ for $x=0$ and  $\alpha=0,~0.5,~1.0$.}
\label{fig1}
\end{figure}

Having in mind that for a given black hole $T_c=r_{+c}/\pi$, 
it can be verified that the critical temperature is given by
\be 
T_c(\alpha) = \frac{\rho^{\frac{1}{3}}}{\pi \Lambda_{min}^{\frac{1}{3}}}~
\bigg( \frac{\tilde{\beta}-\frac{\alpha^2}{4}\frac{\rho_D}{\rho}}{\tilde{\alpha}}\bigg)^{\frac{1}{6}}.
\ee
In the present  model of p-wave metal/superconductor phase transition, we find that $T_c$ depends on
the $\alpha$-coupling constant of {\it dark matter} sector even in the probe limit. 
This dependence reduces to the  $({\tilde \alpha}/{\tilde \beta})^{-1/6}$ form for $\rho_D=0$, which implies that 
the bigger $\alpha$ we take into account, the bigger value of the critical temperature one obtains.
On the contrary, without backreaction, we have no such effect in s-wave holographic metal/superconductor
phase transition in the theory with $U(1)$-gauge {\it dark matter} sector \cite{nak15}.
For general values of dark matter density the behavior is more complicated and is shown in figure \ref{fig1}, for a few values of $\rho_D$. The
value of $T_c(0)= \frac{\rho^{\frac{1}{3}}}{\pi \Lambda_{min}^{\frac{1}{3}}}$ is the
transition temperature of the considered superconductor in the absence of {\it dark matter}.

The previous experience from the study of {\it dark matter} sector effects in the  backreacting 
s-wave holographic superconductor background \cite{nak14}, makes us to believe that the effects should be much stronger
for p-wave symmetry superconductors, but this will be a subject of the future studies.
Evaluating the Sturm-Liouville functional (\ref{SL-l2}) we have found $\lambda_{min}\approx 4.091$  
for $a\approx 0.6845$. For $\rho=1$ one gets $T_c(0) \approx 0.199$.
In the left panel of figure \ref{fig1}, the dependence of the transition temperature on
the coupling to the {\it dark matter} has been shown for a number of values $\rho_D/\rho$. 
We have normalized $T_c$  by dividing it by $T_c(0)$.

\subsection{Condensation value}
In this subsection, we shall pay attention to the influence of the {\it dark matter} sector
on the condensation operator for the phase transition in question and its temperature dependence
for $T$ not far from the critical one, $T_c$. One assumes a small
but finite value of the condensate $<\cR>$ operator and writes
\be
\phi(z)=\frac{\rho}{r_{+c}^2}(1-z^2)+<\cR>\xi(z).
\label{fi-ksi}
\ee
The function $\xi(z)$ corrects the function $\phi(z)$ evaluated at $T_c$ close to $z\approx 0$.
At temperature $T$ the black hole is characterized by the horizon $r_+$ such that $T=r_+/\pi$.
The general behavior of the field $\phi(z)=\frac{\rho}{r_+^2}(1-z^2)$ is again deduced from Eq. (\ref{bc-inf}).
Expanding $\xi(z)$ around $z=0$ 
\be
\xi(z)= \xi(0) + \xi'(0)~z + \frac{1}{2}\xi''(0)~z^2 + \dots ,
\ee  
one deduces from  $\phi(z=1)=0$ that $\xi(1)=0$. Comparing the coefficients
multiplying the same powers  of $z$ in the two expressions for $\phi(z)$ we get
\ben
\frac{\rho}{r_+^2}=\frac{\rho}{r_{+c}^2} +<\cR>\xi(0) \\ \label{gap}
\frac{\rho}{r_+^2}=\frac{\rho}{r_{+c}^2} -\frac{1}{2}<\cR>\xi''(0) \\
\xi'(0)=0
\een 
Consistency of the above equations requires $\xi(0)=-\xi''(0)/2$. To find temperature
dependence of $<\cR>$ we need the value of $\xi''(0)$. 

In order to obtain $\xi''(0)$ we use the equation (\ref{met1}), introduce the expression (\ref{fi-ksi}) 
into it and get the relation for $\xi(z)$, valid  up to the second order in $\cR$ 
\be
\xi'' - \frac{\xi'}{z} =\frac{\rho<\cR>}{\tilde \alpha~r_+^2r_{+c}^2} \frac{z^2~F^2(z)(1-z^2)}{g_0}.
\ee
It can be easily verified, by multiplying the above equation by $z$, 
that $\xi'(0)\equiv 0$.  It also provides  that $\xi''(0)$  is finite. 
To find $\xi''(0)$ we rewrite left-hand side of the last equation as $\p (\xi'(z)/z)/\p z$ and 
 use the boundary conditions $\xi'(1)=0$. It implies 
\be
\xi''(0) = \frac{\xi'(z)}{z} \mid_{z \rightarrow 0} = - \frac{\lambda}{\tilde \alpha}
\int_0^1 dz~\frac{z~(1-z^2)~F^2(z)}{g}=-\frac{\rho<\cR>}{2\tilde \alpha~r_+^2r_{+c}^2}E,
\label{ksi-prim}
\ee
where we set
\be
E = \int_0^1 dz~\frac{z^3~F^2(z)}{(1+z^2)}.
\ee
Making use of Eqs. (\ref{gap}) and (\ref{ksi-prim}) we find the
dependence of the condensation value on temperature close to $T_c$
\be
<\cR> = \sqrt{\frac{2\tilde{\alpha}\pi^2}{E}} T_c(\alpha)\sqrt{1+\frac{T}{T_c(\alpha)}}~\sqrt{1 - \frac{T}{T_c(\alpha)}}.
\ee
Using the previously found value of $a=0.6845$, we obtain $E\approx 0.054046$ and temperature independent prefactor
calculated for $\alpha=0$ is $<\cR>_0 \approx 3.803$.

The right panel  of  figure  \ref{fig1} illustrates the dependence of the
condensation operator $<\cR>$ normalized to its $\alpha=0$ value $<\cR>_0$ on temperature $T$,
normalized to the actual transition temperature  for 
three values of $x=\rho_D/\rho=0,~3,~6$ and  $\alpha=0.5$. The condensation operator is a decreasing function
of $x$, for a given value of $T/T_c(\alpha)$. The dependence of the condensation value on
temperature for $x=0$ and three values of  $\alpha=0,~0.5,~1$ is shown in the inset for $x=0$.  
The value of $<\cR(T)>$  decreases with the  coupling $\alpha$ for constant ratio $T/T_c(\alpha)$ for
all values of $x$. Its behavior as a function of $T$ is shown in the inset to the right panel of
figure  \ref{fig1} for three values of $\alpha$.
 
The above equation envisages the fact that $<\cR>$ depends on $\alpha$-coupling constant of the {\it dark matter} sector
directly {\it via}  factor $\tilde{\alpha}^{1/2}$ and indirectly through $T_c(\alpha)$. It also depends on the
density $\rho_D$ of the dark matter {\it via} $T_c$. The  transition temperature increases with the coupling $|\alpha|$ 
for $\rho_D=0$. It features a monotonous increase with $\alpha$ for $x=1$ and strongly decreases 
with $|\alpha|$ for the value  $\rho_D/\rho=6$, 
close to that inferred from astronomical observations. 
On the AdS/CFT side, the operator $<\cR>$ can be interpreted as responsible for the pairing mechanism. The smaller vacuum
expectation value it has, the harder condensation happens. So we conclude that {\it dark matter} sector 
destructively influences  the condensation phenomena in p-wave superconductors for $\rho_D/\rho>1$. 
As visible from figure  \ref{fig1} different behaviors are expected for $\rho_D/\rho=0.1$. The opposite 
(to that in the case $\rho_D/\rho>1)$ conclusion
was achieved in the case of s-wave holographic metal/superconductor phase transition \cite{nak15a}.
Interestingly, in the Maxwell p-wave model we expect similar dependencies on dark matter sector
as in s-wave superconductors. The observation of the effect in the laboratory could 
shed some light on the meaning of various models of holographic superconductors and their applicability
for the description of real life materials.  
In the case of Gauss-Bonnet p-wave metal/superconductor phase transitions \cite{li11}, the 
curvature corrections also imply the growth of the value of the condensation
operator. This fact was also confirmed by the previous studies \cite{gre09}-\cite{pan10}.

\section{Insulator/ holographic p-wave superconductor phase transition}   
\label{sec:insulator sup}
The main ingredient in the considerations of insulator/superconductor phase transition will be
the gravitational background of five-dimensional AdS soliton spacetime, given by
\be
ds^2 = -r^2~dt^2 + L^2~\frac{dr^2}{f(r)} + f(r)~d \varphi^2 + r^2~(dx^2 + dy^2),
\label{sol}
\ee
where $f(r)= r^2 - r_0^4/r^2$,~$r_0$ denotes the tip of the line element which constitutes a conical singularity
of the considered solution. Without loss of generality we set the radius of the AdS spacetime equal to one.
The AdS solitonic solution can be gained from the five-dimensional Schwarzschild-AdS black hole line element
by making two Wick rotations. On the other hand, a conical singularity, at the tip $r=r_0$, can be get rid of
by the Scherk-Schwarz transformation of $\varphi$-coordinate in the form $\varphi \sim \varphi + \pi/r_0$.
The gravitational background in question delivers the description of a three-dimensional field theory
with a mass gap which in turn takes after an insulator in the condensed matter physics. The temperature
of the aforementioned background equals to zero.

In order to solve the underlying equations of motion for the p-wave holographic insulator/superconductor
phase transition problem, we shall impose the adequate boundary conditions. Namely, at the tip one has the same
boundary conditions as in the s-wave insulator/ superconductor problem
\ben
w &=& w_0 + w_1~(r-r_0) + w_2~(r-r_0)^2 + \dots,\\
\phi &=& \phi_0 + \phi_1~(r-r_0) + \phi_2~(r-r_0)^2 + \dots,
\een
where $w_a$ and $\phi_a$, for $a = 0,~1,~2, \dots$ are  constants. One encumbers the Neumann-like
boundary condition to obtain every physical quantity finite \cite{nis10}. Contrary, near the boundary
where $r \rightarrow \infty$, we have the different asymptotical behavior (comparing to the s-wave case).
The asymptotic solutions read 
\be
w \rightarrow w_0 + \frac{w_2}{r^2}, \qquad \phi \rightarrow \mu - \frac{\rho}{r^2},
\ee
where $\mu$ and $\rho$ are interpreted as the chemical potential and the charge density in the dual theory, respectively.
On the other hand, $w_0$ and $w_2$ have interpretations as a source and the expectation value of the dual operator.
In order to gain the normalizable solution, one puts $w_0 =0$ (we are not interested in the case when
the dual operator is  sourced).
  
In $z$-coordinates (with $r_0=1$) the equations in question yield
\ben \label{w1}
w''(z) &+& \bigg( \frac{f'(z)}{f(z)} + \frac{1}{z} \bigg)~w'(z) 
+ \frac{\phi (z)[\phi(z)-\frac{\alpha^2}{4}~\eta(z)]}{\tilde \alpha~
f(z)~z^2}~w(z)= 0,\\ \label{w2}
\phi''(z) &+& \bigg( \frac{f'(z)}{f(z)} + \frac{1}{z} \bigg)~\phi'(z) - \frac{w^2(z)}{\tilde \alpha~z^2~f(z)}~\phi(z)
= 0.
\een

\subsection{Critical chemical potential}
It was revealed numerically \cite{nis10} that when the chemical potential exceeds its critical value 
the condensation operator turns on. In the case when
$\mu < \mu_c$ the scalar field $w(r)$ is equal to zero, which is interpreted as the insulator phase.
The system under consideration has a mass gap bounded with the confinement in $(2+1)$-gauge 
theory via the Scherk-Schwarz compactification. It 
justifies the conclusion that the critical value of the allowed potential is the turning point 
of the insulator/superconductor phase transition.
Moreover, when $\mu \approx \mu_c$, the scalar field $w(r)$ is so small that $w^2(r)=0$ and 
the equation (\ref{w2}) has an analytical 
solution in the form of a logarithmic function $\phi(z)=C_1+C_2[\log(1+z^2)-\log(1-z^2)]$ ~\cite{cai11b}.
Establishing the value of the integration constant at the tip $z=1$ ($C_2=0$), one gains that $\phi(z)$ 
has the constant value $C_1=\mu$. 
From the boundary conditions, it can also be seen that near $z=0$ we get $\rho =0$.

As in the preceding section we find the dependence of $\phi (z)$ on the component of the {\it dark matter} sector $\eta(z)$.
Using the the adequate components of the metric tensor for the line element (\ref{sol}) and equation (\ref{eta}), 
 we arrive at the following relation  
\be
\eta(z) + \frac{\alpha}{2}~\phi(z) = D_1~\bigg[ \frac{z^2}{2}  + \frac{1}{4} \log \bigg( \frac{1+z^2}{1-z^2} \bigg) \bigg]+ D_2,
\ee
where $D_1$ and $D_2$ are integration constants. Setting the integration constant 
$D_1$ equal to zero at the tip $z=1$, we obtain 
\be
D_2 = \mu_D + \frac{\alpha}{2}~\mu,
\ee
where $\mu_D$ denotes the chemical potential of the {\it dark matter}. 
For $\mu$ tending to  its critical values $\mu_c$ from above, the equation of motion for $w(z)$ function is of the form
\be
w''(z) + \bigg( \frac{f'(z)}{f(z)} + \frac{1}{z} \bigg)~w'(z) + \frac{\mu_c[
\mu_c - \frac{\alpha^2}{4}(\mu_D - \frac{\alpha}{2}~\mu_c)]}{f(z)~z^2~\tilde \alpha}~w(z) = 0.
\label{crit}
\ee
To proceed we assume $w_0=$ and correct the solution for $w(z)$ close to the boundary $z=0$
by  defining the trial function $G(z)$
\be
w(z) \sim <\cO>~z^2~G(z),
\ee
with the boundary conditions $G(0)=1$ and $G'(0)=0$ and $<\cO>=w_2$. For simplicity we set $r_0=1$. It implies 
further, that the equation (\ref{crit}) can be rewritten as
\be
(P(z)~G'(z))' - Q(z)~G(z) + {\tilde \mu}^2~R(z)~F(z) = 0,
\ee
where we have defined the quantities as follows:
\ben
\tilde{\mu}^2 &=& \frac{\mu_c^2}{\tilde{\alpha}}~\left(\tilde \beta - \frac{\alpha^2}{4}~\frac{\mu_D}{\mu_c}\right),\\
P(z) &=& z^5~f,\\
Q(z) &=& -f~\bigg( 4~z^3 + 2~z^4~\frac{f'}{f} \bigg),\\
R(z) &=& {z^3}.
\een
The minimum eigenvalue of $\tilde{\mu}^2$ can be achieved from the variation of the functional
\be
{\tilde \mu}^2 = 
\frac{\int_0^1 dz~[G'(z)^2~P(z) + Q(z)~G^2(z)]}{\int_0^1 dz~R(z)~G^2(z)}.
\ee
With the choice of  $G(z) = 1 - a~z^2$ as the trial function we find $\tilde{\mu}_{min}\approx 2.2666$ 
for $a\approx 0.3376$. Let's note that for $\alpha=0$ we have $\mu_c^2=\tilde{\mu}^2$ as it should be. The
dependence of $\mu_c$ on $\alpha$ requires knowledge of $\mu_D$, which is unknown.
For illustrational purposes we shall calculate the dependence $\mu(\alpha)$ for two  values of $\mu_D=0, ~\mu_c$.
In the first case we end up with the dependence $\mu=\tilde{\mu}_{min}\sqrt{\tilde{\alpha}/\tilde{\beta}}$,
while in the second case it has the form  $\mu=\tilde{\mu}_{min}\sqrt{\tilde{\alpha}/[\tilde{\beta}-\alpha^2/4]}$.

The dependence of the critical chemical potential on $\alpha$ is shown in the left panel of figure  \ref{fig2}, for both values of $\mu_D$. For $\mu_D=0$ the 
chemical potential is the decreasing function of
$\alpha$, independently of its sign. Contrary to that for $\mu_D=\mu_c$ the critical chemical
potential increases for negative values of the coupling to dark matter, while decreases for positive $\alpha$.
Thus, the transition appears at lower values of the chemical potential $\mu_c$ for positive values of $\alpha$,
independently if $\mu_D$ vanishes or equals that of normal matter.
This confirms the fact that {\it dark matter}  makes the p-wave phase transition easier to happen
provided the coupling is positive.
Let's recall that for  s-wave symmetry the holographic insulator/superconductor phase transition 
 the chemical potential was  unaffected by the presence of {\it dark matter} sector \cite{nak15a}. The result was  
similar as in Einstein-Maxwell theory \cite{nis10,cai11c}. 

On the contrary, in the Gauss-Bonnet gravity the critical potential increases with the growth of the 
curvature corrections \cite{pan11}, making the condensation harder to form. This phenomenon
was also found in s-wave holographic insulator/superconductor phase transition in Gauss-Bonnet system.

\subsection{Critical phenomena}
When $\mu \rightarrow \mu_c$ the condensate value $<\cO>$ is small but finite and the equation 
of motion for the $\phi$ field is provided by
\be
\phi'' + \bigg( \frac{f'}{f} + \frac{1}{z} \bigg)~\phi' - 
\frac{<\cO>^2~z^2~F^2(z)}{\tilde \alpha~f}~\phi=0.
\label{cr0}
\ee
For $\mu$ slightly above the critical value the condensation scalar operator $<\cO>$ 
is very small. This enables us to look for the solution in the form 
\be
\phi(z) \sim \mu_c ~+~ <\cO>~\chi(z) + \dots
\label{cr1}
\ee
Moreover, to recover the previous result $\phi(z)=\mu$, one imposes the boundary condition $\chi(1)=0$
at the tip of the considered soliton.
Near the boundary $z=0$, one expands the function 
$\chi(z)=\chi(0)+\chi'(0)z+\frac{1}{2}\chi''(0)z^2+\ldots$, and rewrites $\phi$ in the form given by
\be
\phi(z) \simeq \mu - \rho~z^2 \simeq
\mu_c + <\cO>\bigg( \chi(0) + \chi'(0)z + \frac{1}{2}\chi''(0)~z^2 + \dots \bigg),
\label{expan}
\ee
where we have imposed the boundary conditions at the tip of the soliton for the function $\chi(z)$ requiring that
$\chi(1) = 0$. Comparing the coefficients of the $z^0$-terms, in the above equation, we find that
\be
\mu - \mu_c \simeq~ <\cO> \chi(0).
\ee
Having in mind the relations (\ref{cr1}) and (\ref{cr0}), one can easily find the relation
for $\chi(z)$. Namely, it is given by 
\be
\chi''(z) + \bigg( \frac{f'}{f} + \frac{1}{z} \bigg)~\chi'(z) - 
\frac{<\cO>~\mu_c~z^2~F^2(z)}{\tilde \alpha~f} + O(<\cO>^{n\geq 2}) = 0,
\ee
where by $O (<\cO>^{n\geq 2})$ we denoted terms of order $n \geq 2$ which can be neglected
because of the fact that they are significantly smaller than the linear one.
By virtue of the above we arrive at the  relation
\be
\chi''(z) + 
\bigg( \frac{f'}{f} + \frac{1}{z} \bigg)~\chi'(z) - 
\frac{<\cO>~\mu_c~z^2~F^2(z)}{\tilde \alpha~f} \simeq 0.
\label{cr2}
\ee
Let us redefine $\chi(z)$ function by the new one $\xi(z)$ 
\be
\chi(z) = \frac{<\cO>~\mu_c }{ {\tilde{\alpha}}} ~\xi(z).
\ee
The new definition of $\chi(z)$ allows to get rid of the ${\tilde{\alpha}}$-coupling dependence
in the relation (\ref{cr2}). One obtains 
\be
\xi'' + \bigg( \frac{f'}{f} + \frac{1}{z} \bigg)~\xi' 
- \frac{z^2~F^2(z)}{f}= 0.
\ee
Consequently, the scalar operator $<\cO>$ provides the following:
\be
<\cO> = \sqrt{\frac{(\mu - \mu_c)~\tilde{\alpha}}{\mu_c~\xi(0)}},
\label{scalop}
\ee
where $\xi(0)$ is given by 
\be
\xi(0)= a_1 - \int_0^1 \frac{dz}{ f~z}~\bigg(a_2 + \int_1^z dy ~
y^3~F^2(y) \bigg).
\label{xi0}
\ee
The integration constants $a_1$ and $a_2$, are determined by the boundary conditions imposed on $\chi(z)$-function.
To find them we have to fulfill the requirement $\xi(1)=0$. This leads to $a_2=0$ required to cancel the
$a_2\log(1-z^2)$ term resulting from the integration over $z$ 
and  $a_1=(3-2a+2a^2-4a\log2)/24$. Using the previously found optimal value of $a=0.3375$ one ends 
up with $\xi(0)=a_1\approx 0.0894$.

On the other hand, the above relations reveal that the operator $<\cO>$ yields
\be
<\cO> \simeq ~\Delta~(\mu - \mu_c)^\frac{1}{2},
\label{opsc}
\ee
where the prefactor 
$\Delta=\sqrt{\frac{\tilde{\alpha}}{\mu_c\xi(0)}}\approx 3,139\{\tilde{\alpha}[\tilde{\beta}-(\alpha^2/4)(\mu_D/\mu)]\}^{1/4}$  
contains information on the dependence on {\it dark matter} sector. The previously found relation  
 $\mu_c(\tilde{\alpha})=\tilde{\mu}\sqrt{\tilde{\alpha}/[\tilde{\beta}-(\alpha^2/4)(\mu_D/\mu)]}$ 
has to be taken into account. 

Equation (\ref{opsc}) envisages  that the p-wave holographic/superconductor
phase transition represents the second order phase transition for which the critical exponent
has the mean field value $1/2$. This was also the case in s-wave insulator/superconductor phase transition
influenced by the {\it dark matter} sector \cite{nak15}. Thus the {\it dark matter} sector does not change the
order of the transition.

\begin{figure}
\includegraphics[width=0.5\linewidth]{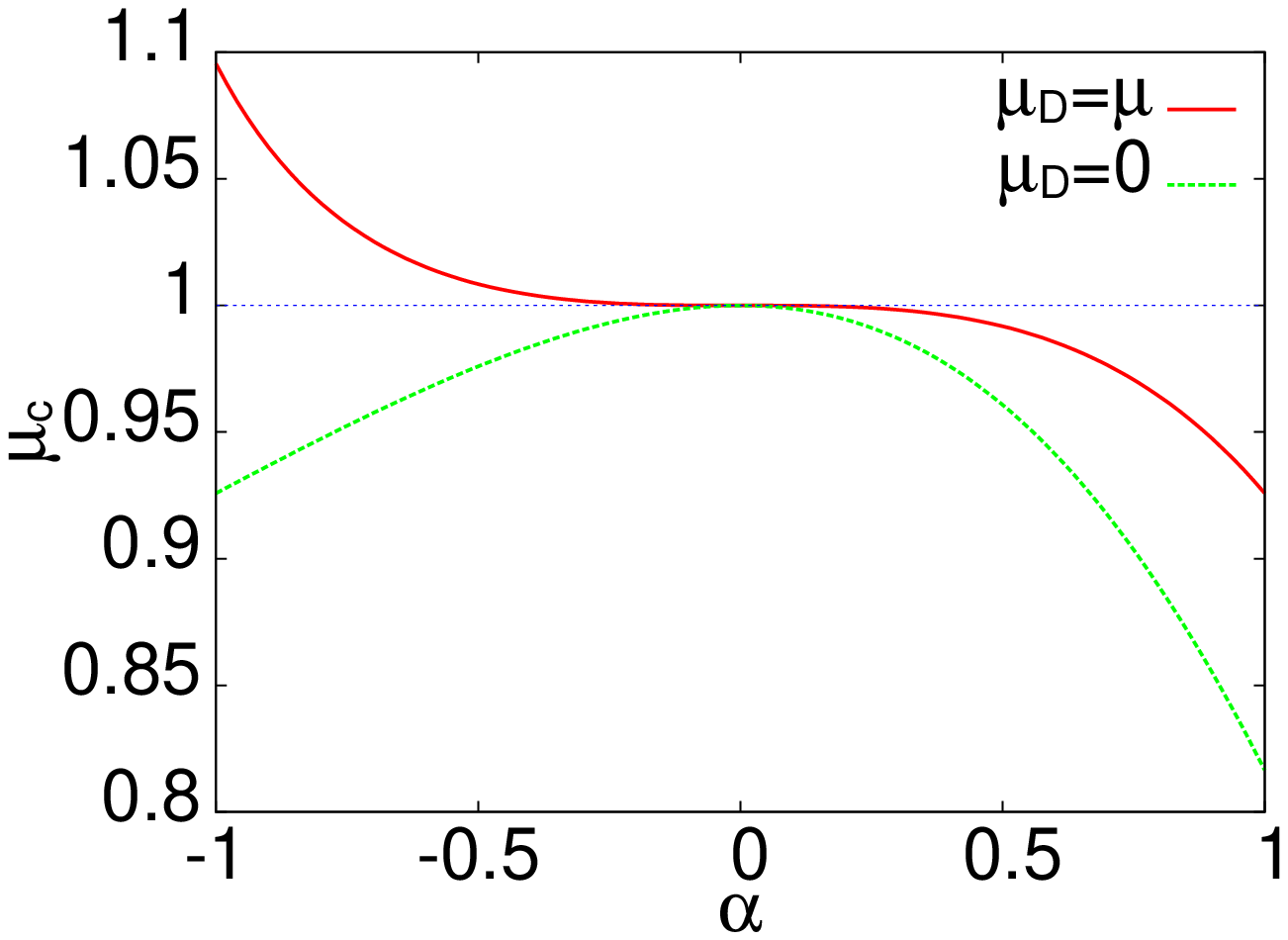}
\includegraphics[width=0.5\linewidth]{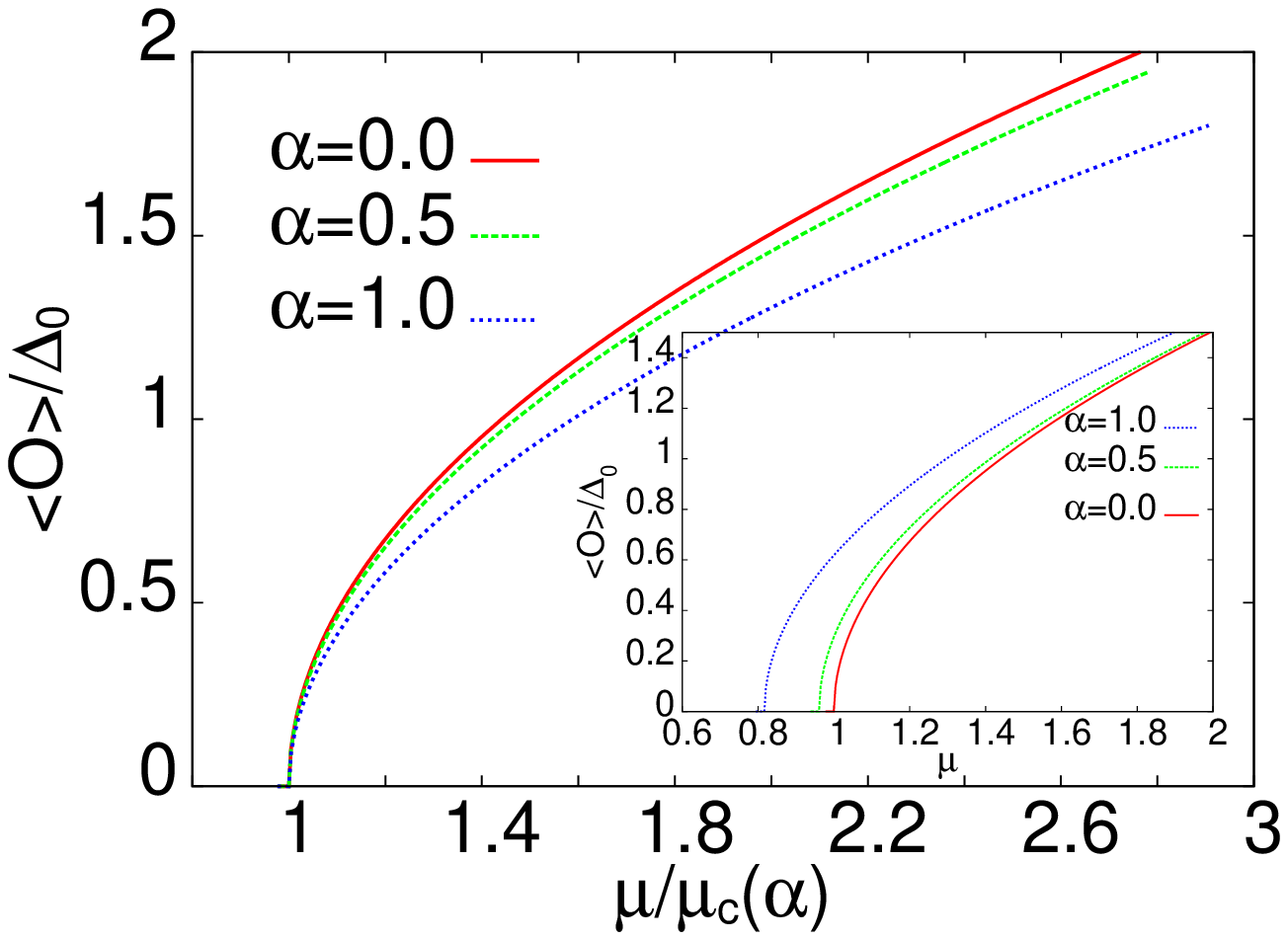}
\caption{(color online)  
Left panel shows the dependence of the critical value of the chemical potential $\mu_c$, measured
in units of $\tilde{\mu}$, for the 
insulator - holographic p-wave superconductor transition {\it vs.} $\alpha$ for two values of the
dark matter chemical potential $\mu_d=0,\mu_c$.
The dependence of the condensation value  $<\cO>$ normalized to its $\alpha=0$ amplitude $\Delta_0$ 
on the normalized chemical potential $\mu/\mu_c(\alpha)$ is shown for $\mu_D=0$ and a few values
of the coupling to the {\it dark matter} sector $\alpha=0, 0.5, 1.0$. The inset
to the right panel shows similar dependence on the bar value of $\mu$, and for $\mu_D=0$.}
\label{fig2}
\end{figure}

The right panel of figure \ref{fig2} illustrates the dependence of the condensation value $<\cO>$ on $\mu$ close to $\mu_c$.
The dependence on $\mu/\mu_c(\alpha)$ shows that the amplitude of $<\cO>$ diminishes with $\alpha$. In fact 
the dark matter increases the value of the condensate operator $<\cO>$ if $\mu$ is kept constant as is visible 
from the inset. This conclusion is independent of the value of $\mu_D$ provided $\alpha>0$.

Next, we proceed to find the dependence of the charge density $\rho$ on the critical chemical potential.
In order to calculate charge density $\rho$, we consider $z^1$-order coefficients in the relation
(\ref{expan}), having in mind that
$\xi'(0)=0$, together with the previous requirement $\xi(1)=0$ being subject to the boundary condition.
It leads to 
\be
\chi''(0)={\chi'(z) \over z} \mid_{z \rightarrow 0} = - <\cO>~\frac{\mu_c}{\tilde{\alpha}}
~\int_0^1 dz~z^{3}~F^2(z).
\label{li}
\ee
The comparison of the adequate coefficients of $z^2$-order in equation (\ref{expan}) leads to the conclusion that
the charge density implies
\be
\rho = - {<\cO> \over 2}~\chi''(0).
\label{denc}
\ee
By virtue of the equations (\ref{li}) and (\ref{denc}), 
having in mind the relation (\ref{scalop}), one arrives at the following
\be
\rho = (\mu - \mu_c)~{\tilde{ D}},
\ee
where the quantity $\tilde{ D}$ yields
\be
\tilde{ D} = {1 \over 2~\xi(0)}~\int_0^1 dz~z^{3}~F^2(z).
\ee 
Evaluating the integral with the previously found parameters and introducing $\xi(0)$, we obtain $\tilde{D}\approx 0.844$.

\section{P-wave holographic droplet}
\label{sec:droplet}
In this section, based on the notion of marginally stable modes of vector perturbations
we shall investigate the influence of the magnetic field on holographic p-wave insulator/superconductor
phase transition. We restrict the considerations to the probe limit.

To begin with, let us recall that the stability of a spacetime can be studied by the
quasinormal modes (QNMs) of the perturbations in the background in question. When one encounters
that the imaginary part of QNMs is negative, then they will decrease with the passage of time and
disappear. The conclusion is that the elaborated spacetime is stable against these perturbations.
On the other hand, the positivity of the imaginary part of QNMs reveals that the spacetime is
unstable against the perturbations.
The signal of instability or possible phase transition in the spacetime in question, is the occurrence
of marginally stable modes ($\omega = 0$). They are only dependent on radial coordinates and do not
backreact on the other fields.

As the gravitational background
we shall use the line element of the five-dimensional AdS soliton,
rewritten in the coordinates $(t,~r,~\tilde{\rho},~\varphi,~\theta)$. It has the form
\be
ds^2 = - r^2~dt^2 + {dr^2 \over f(r)} + f(r)~d\varphi^2  + r^2 ~(d\tilde {\rho}^2 + \tilde{\rho}^2 ~d\theta^2),
\ee
with $f(r)=r^2-r_0^4/r^2$.
We assume the following components of the gauge fields $A_{\mu}^{(b)}$ and
$B_{\mu}^{(b)}$:
\ben 
A &=& (\mu_c~dt + \frac{1}{2}B_f~\tr^2~d \theta)~\tau^3 + \psi(r,~t,~\varphi,~\tr)~\tau^1~d \rho,\\
B &=& \mu_D~\tau^3~dt,
\een
where $B_f$ is the constant magnetic field. The time components of both fields $\mu_c$ and $\mu_D$ have 
constant values. This ansatz is valid close to the critical point of the phase transition. As in a previous section
$\mu_c$ is the critical chemical potential for the phase transition in a droplet, and $\mu_D$ plays the role
of the chemical potential of a dark matter.

The $\rho(1)$ component of the underlying equations of motion is provided by 
\be
\p^2_r \psi + \bigg( {\p_r f \over f} + {1 \over r} \bigg)\p_r \psi + {1 \over f^2} \p^2_{\varphi} \psi
- {1 \over f} \p^2_t \psi  +
{1 \over r^2~f ~\tilde \alpha}~\bigg[
\mu_c~\bigg(\mu_c -\frac{\alpha^2}{4}~\mu_D \bigg) - \frac{B_f^2~\tr^2}{4} \bigg]~\psi = 0.
\ee
In order to solve the above equation we choose an ansatz for $\psi$ field which implies
\be
\psi = F(r,~t)~H(\varphi)~U(\tr).
\ee
We set $U(\tr) = \tr = 2/\sqrt{B_f}$, because of the fact that one looks for the solution confined to a finite
circular region which radius is proportional to $1/\sqrt{B_f}$. Consequently, we arrive at the relations 
\ben
\p^2_r F &+& \bigg( {1 \over r} + {\p_r f \over f}\bigg)~\p_r F - { r \over f}~\p^2_t F
+
\frac{1}{r^2~f}~\bigg( \frac{\mu_c~(\mu_c - \frac{\alpha^2}{4}~\mu_D)}{\tilde \alpha}
- \frac{\la^2~r^2}{ f} - \frac{B_f}{\tilde \alpha} \bigg)~F = 0,\\
\frac{\p^2_{\varphi} H}{ H } &=& - \la^2.
\een
One considers the case when $\la = 2r_0n/L$, where $n \in Z$ causing the periodicity
in $H(\varphi) = H(\varphi + \pi L/r_0)$.
In what follows without loss of the generality
we set $r_0 =1$ and $L=1$. Just, from periodicity property of $H(\varphi)$ we identify $\la = 2n$.

In the next step,
one substitutes $F(r,~t) = e^{-i \omega t}~R(r)$. Consequently, the requirement 
concerning marginally stable modes leads to the condition that $\omega =0$.
Next, the redefinition of the $r$-coordinates in terms of $z$ ones, provides the relation for
$R(z)$
\be
\p^2_z R(z) + \bigg( {\p_z f \over f} + {1 \over z} \bigg)~\p_z R(z) 
+ {1 \over z^2~f}~\bigg(
\frac{\mu_c~(\mu_c - \frac{\alpha^2}{4}~\mu_D) - B_f}{\tilde \alpha} - \frac{4~n^2}{z^2~f} \bigg)~R(z) = 0. 
\ee
Then, one introduces a trial function in the form
\be
R(z) \mid_{z \rightarrow 0} \sim~ <\cO>~z^{2}~\Theta(z),
\ee
with the boundary conditions imposed on $\Theta(z)$. Namely, $\Theta(0) =1$ and $\Theta'(0) = 0$.\\
After some algebra, the resulting equation can be 
converted into the standard Sturm-Liouville eigenvalue equation, which can be written as
\be
\p_z\bigg( a(z)~\Theta' \bigg) - b_n(z)~\Theta + \delta_n^2~c(z)~\Theta = 0,
\ee
where we have denoted by $\delta_n^2$ the relation
\be
\delta_n^2 = \frac{\mu_c~(\mu_c - \frac{\alpha^2}{4}~\mu_D) - B_f}{\tilde \alpha}.
\ee     
The subscript $n$ is used to remind that the function $b(z)$ depends on the parameter $n$.  
                                            
The remaining quantities are defined by
\ben
a(z) &=& f~z^{5},\\
b_n(z) &=& - f~z^{4}~
\bigg( {2 \over z} + 2~\bigg( {\p_z f \over f} + {1 \over z} \bigg) - \frac{4~n^2}{z^3~f^2}\bigg),\\
c(z) &=& z^{3}.
\een
The eigenvalues of $\delta_n^2$ can be found by the method of minimizing  the functional in question
\be
\delta^2 = \frac{\mu_c~(\mu_c - \frac{\alpha^2}{4}~\mu_D) - B_f}{\tilde \alpha}
= {\int_0^1 dz~(\Theta'(z)^2 ~a(z) + b_n(z)~\Theta(z)^2)
\over \int_0^1 dz~c(z)~\Theta^2(z)}.
\label{muc-dropl}
\ee

To minimize the above functional we have chosen as the trial function $B(z) = 1 - a z^2$. The Sturm - Liouville
minimization gives $a\approx 0.942$ and $\delta^2\approx 3.8279$ for $n=1$, while
one gets $a\approx 1.0083$ and $\delta^2\approx 5.6034$ for $n=2$.
It can be seen that not only magnetic field influences the condensation but also $\alpha$ coupling constant 
of the {\it dark matter} sector as well as the component of the {\it dark matter} gauge field $B_t^{(3)}$, i.e., its
chemical potential. 

\begin{figure}[h]
\centerline{
\includegraphics[width=0.6\linewidth]{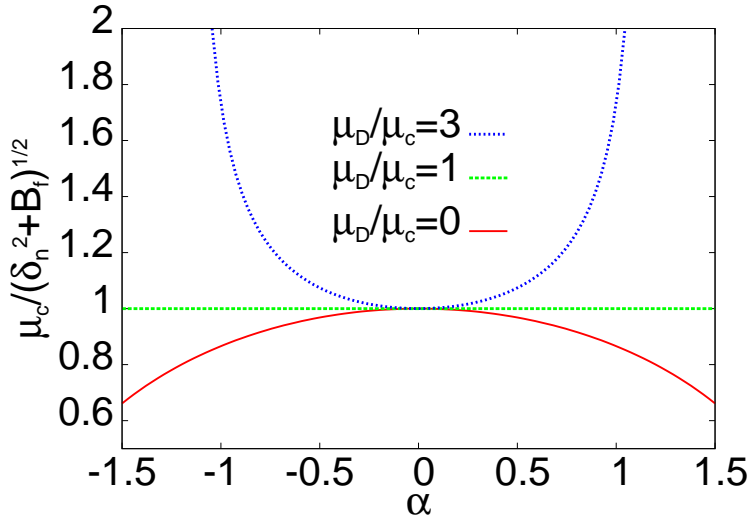}}
\caption{(color online). The dependence of the chemical potential $\mu_c$ normalized to $\sqrt{\delta_n^2+B_f}$ 
on the coupling constant $\alpha$ for the p-wave droplet for a few values of the parameter $x=\mu_D/\mu_c$.}
\label{fig3}
\end{figure}

It can be noted that for $\mu_D=\mu_c$ the dependence on $\alpha$ cancels out. 
Having found $\delta_n^2$ we get the formula for $\mu_c$ the critical value of the chemical
potential for metal-superconductor phase transition at temperature zero. Denoting the ratio of chemical potentials by
$x=\mu_D/\mu_c$, one gets
\be
\mu_c=\sqrt{\delta_n^2+B_f}\left(\frac{\tilde{\alpha}}{1-x ~\alpha^2/4}\right)^{1/2}.
\label{muc-dropl2}
\ee
The dependence of the chemical potential $\mu_c$ normalized by its $\alpha=0$ value equal to $\sqrt{\delta_n^2+B_f}$ 
on the coupling constant $\alpha$ is shown in the figure  \ref{fig3}, for three values of the {\it dark matter} chemical potential.
For illustrational purposes we have chosen $\mu_D=0,~\mu_c$ and $3\mu_c$. For $\mu_D= 0$ the 
critical value of the chemical potential
$\mu_c$ is a decreasing function of $|\alpha|$. It takes on constant value for $x=1$ and is strongly increasing function of
 $|\alpha|$ for $x>1$. 

\section{Summary and conclusions}
\label{sec:conclusions}

In the paper we have analytically investigated phase transitions towards p-wave holographic superconductors
in which {\it dar matter} sector can imprints its presence.
All the calculations were conducted in the probe limit, when the gravitational background
was described by the five-dimensional AdS Schwarzschild black hole  metric and AdS solitonic 
line element, respectively.

It was revealed that by a special choice of the real vector field, the vector Maxwell 
model of p-wave holographic superconductor is equivalent to s-wave description elaborated in \cite{nak15a}.
All the properties of the considered phase transitions, like insulator/ superconductor and metal/superconductor phase transitions,
in both models look identically. Of course, one should have in mind that the s-wave order parameter is replaced
by $\rho_x$ component of the vector field, being in p-wave holographic superconductor an order parameter.
Due to that fact we have considered in detail the $SU(2)$ version of p-wave holographic superconductor.
Our ansatz for $SU(2)$-gauge fields includes two $U(1)$ groups, the  subgroups of $SU(2)$, one describing
the ordinary Maxwell field and the other related to the {\it dark matter} sector.

Using the AdS/CFT approach and studying the properties of holographic superconductors
in the background with {\it dark matter} sector we rely on the following three convictions.
Firstly, we accept that the gauge-gravity duality teaches us about strongly coupled 
superconductors which are produced and studied in the laboratories. Some examples
of this can be found in \cite{gre13}. Secondly, the AdS/CFT correspondence relies on the AdS spacetime, while
our space time is rather the dS  one. The power of the gauge/gravity duality  
does not depend  whether the Universe in which 
the superconductors exist, itself forms the  AdS or de Sitter  spacetime. It serves as a kind of calculus, 
which enables tackling strong coupling problems in a $d$ - dimensional field theory using perturbative 
approach in $d+1$ - dimensional gravity theory.
Third, the dark matter existence seem to be obvious from astrophysical data as discussed
in the Introduction. If it interacts with ordinary matter as 
proposed  in  the relation (\ref{s_matter}) and quantified by the coupling constant $\alpha$, 
so it modifies the behavior of the ordinary matter as visible from the equations (\ref{e4}). These
modifications are found to influence the superconducting transition temperature and chemical potential 
and could, in principle, be observed as the annual changes of the properties
of superconductors  following the expected annual changes in the distribution of the
dark matter \cite{freese2013}.

Studies of the p-wave holographic metal/superconductor phase transitions reveal the critical temperature
dependence on $\alpha$ - the coupling constant of the {\it dark matter} sector and other parameters of the
dark matter: its chemical potential $\mu_D$ and its density $\rho_D$ - both unknown. So for illustrational purposes
we have assumed that they may take arbitrary, albeit physically sensible values. 
The results depend on these values. For example 
it turns out that for $\rho_D=0$ the greater value of $\alpha$ one sets, the bigger value of the critical temperature
one receives, $c.f.$ the left panel in figure \ref{fig1}. On the other hand for $\rho_D/\rho>1$ the increase of
$\alpha$ results in decrease of $T_c$.
This is an interesting result, which could potentially be of interest in possible experiments aimed at the 
detection of {\it dark matter}. However, one should remark that, the qualitatively similar dependence 
(as here for $\rho_D=0$) was earlier found in the case of the backreacting s-wave holographic superconductor
in the theory under inspection \cite{nak14}. It has to be checked to which extend the effect will be
modified if the backreaction is taken into account in the model under consideration. 
The condensation value $<\cR>$ also exhibits dependence on the  {\it dark matter} coupling constant. 

As far as the holographic insulator/p-wave superconductor phase transition is concerned, it was spotted
that the critical chemical potential also reveals the dependence on $\alpha$.
For $x=0$ one observes the decrease of $\mu_c$ with $|\alpha|$ as shown in figure  \ref{fig2}. In the case of Gauss-Bonnet gravitational background the situation is quite different. 
Namely, the critical potential increases with the development of the curvature corrections. 

The determination of the condensation  operator leads to the conclusion that p-wave holographic insulator/
superconductor  phase transition is of the second order. The order parameter is proportional to $(\mu - \mu_c)^{1/2}$  
with the mean field like exponent. Both $\mu_c$ and the prefactor
depend on $\alpha$. Due to this fact, the condensation value $<\cO>$ may be increasing or decreasing function
of $\alpha$ depending how one measures the distance from the critical chemical potential. If the distance is measured
from the actual $\alpha$ dependent $\mu_c$ (main picture in the right panel of figure  \ref{fig2}) $<\cO>$ is 
decreasing function of $\alpha$. On the other hand, for a given value of $\mu$, the increase of the coupling
to {\it dark matter}  appreciably enlarges the value of the condensation operator. 
On the other hand, the charge density is independent on $\alpha$-corrections except $via$ $\mu_c$.

We have also elaborated p-wave holographic droplet in a constant magnetic field. In our
studies we have used marginally stable modes of vector perturbations. It has been shown that
the critical chemical potential of the superconducting droplet depends on  the {\it dark matter} coupling $\alpha$, 
the magnetic field $B_f$ and  {\it dark matter} chemical  potential $\mu_D$. 
For $\mu_D=0$, $\mu_c$ diminishes as $|\alpha|$ grows. This is similar 
to  the s-wave holographic droplet case with {\it dark matter} sector. For $\mu_D=\mu_c$ there is no dependence
on $\alpha$, while the increase is observed for $\mu_D>\mu_c$.

It is worth mentioning that due to the fact that cosmic as well as on Earth experiments of detecting
{\it dark matter} are in progress, the untypical behavior of $SU(2)$ p-wave holographic superconductors  
during phase transitions in question may give a hint to unveil its secret.

\acknowledgments
We would like to thank the Referee for pointing out the
inconsistency in our earlier choices of the fields $A_\mu$ and $B_\mu$. 
MR was partially supported by the grant of the National Science Center 
$DEC-2013/09/B/ST2/03455$ and KIW by the grant DEC-2014/13/B/ST3/04451.



\begin{thebibliography}{99}

%
\def\cmp#1#2#3#4{\emph{#4}, \emph{ Commun. Math. Phys.} {\bf #1} (#3) #2}
\def\lmp#1#2#3#4{\emph{#4}, \emph{ Lett. Math. Phys.} {\bf #1} (#3) #2}
\def\hpa#1#2#3#4{\emph{#4}, \emph{ Hell. Phys. Acta} {\bf #1} (#3) #2}
\def\grg#1#2#3#4{\emph{#4}, \emph{ Gen. Rel. Grav.} {\bf #1} (#3) #2}
\def\pr#1#2#3#4{\emph{#4}, \emph{ Phys. Rev.} {\bf #1} (#3) #2}
\def\prl#1#2#3#4{\emph{#4}, \emph{ Phys. Rev. Lett.} {\bf #1} (#3) #2}
\def\prd#1#2#3#4{\emph{#4}, \emph{ Phys. Rev. D} {\bf #1} (#3) #2}
\def\pl#1#2#3#4{\emph{#4}, \emph{ Phys. Lett.} {\bf #1} (#3) #2}
\def\pla#1#2#3#4{\emph{#4}, \emph{ Phys. Lett. A} {\bf #1} (#3) #2}
\def\plb#1#2#3#4{\emph{#4}, \emph{ Phys. Lett. B} {\bf #1} (#3) #2}
\def\prep#1#2#3#4{\emph{#4}, \emph{ Phys. Reports} {\bf #1} (#3) #2}
\def\phys#1#2#3#4{\emph{#4}, \emph{ Physica} {\bf #1} (#3) #2}
\def\jcp#1#2#3#4{\emph{#4}, \emph{ J. Comput. Phys.} {\bf #1} (#3) #2}
\def\jmp#1#2#3#4{\emph{#4}, \emph{ J. Math. Phys.} {\bf #1} (#3) #2}
\def\jpm#1#2#3#4{\emph{#4}, \emph{ J. Phys. A: Math. Gen.} {\bf #1} (#3) #2}
\def\cpr#1#2#3#4{\emph{#4}, \emph{ Computer Phys. Rept.} {\bf #1} (#3) #2}
\def\cqg#1#2#3#4{\emph{#4}, \emph{ Class. Quant. Grav.} {\bf #1} (#3) #2}
\def\cma#1#2#3#4{\emph{#4}, \emph{ Computers Math. Applic.} {\bf #1} (#3) #2}
\def\mc#1#2#3#4{\emph{#4}, \emph{ Math. Compt.} {\bf #1} (#3) #2}
\def\apj#1#2#3#4{\emph{#4}, \emph{ Astrophys. J.} {\bf #1} (#3) #2}
\def\apjs#1#2#3#4{\emph{#4}, \emph{ Astrophys. J. Suppl.} {\bf #1} (#3) #2}
\def\acta#1#2#3#4{\emph{#4}, \emph{ Acta Astronomica} {\bf #1} (#3) #2}
\def\apl#1#2#3#4{\emph{#4}, \emph{ Ann. Physik. (Leipzig)} {\bf #1} (#3) #2}
\def\amjp#1#2#3#4{\emph{#4}, \emph{Am. J. Phys.} {\bf #1} (#3) #2}
\def\anp#1#2#3#4{\emph{#4}, \emph{ Ann. Phys.} {\bf #1} (#3) #2}
\def\sa#1#2#3#4{\emph{#4}, \emph{ Sov. Astro.} {\bf #1} (#3) #2}
\def\sia#1#2#3#4{\emph{#4}, \emph{ SIAM J. Sci. Statist. Comput.} {\bf #1} (#3) #2}
\def\aa#1#2#3#4{\emph{#4}, \emph{ Astron. Astrophys.} {\bf #1} (#3) #2}
\def\mnras#1#2#3#4{\emph{#4}, \emph{ Mon. Not. R. Astr. Soc.} {\bf #1} (#3) #2}
\def\npb#1#2#3#4{\emph{#4}, \emph{ Nucl. Phys. B} {\bf #1} (#3) #2}
\def\prsla#1#2#3#4{\emph{#4}, \emph{ Proc. R. Soc. London, Ser. A} {\bf #1} (#3) #2}
\def\jhep#1#2#3#4{\emph{#4}, \emph{ JHEP} {\bf #1} (#2) #3}
\def\nuc#1#2#3#4{\emph{#4}, \emph{ Nuovo Cimento B } {\bf #1} (#3) #2}
\def\ijmp#1#2#3#4{\emph{#4}, \emph{ Int. J. Mod. Phys. D} {\bf #1} (#3) #2}
\def\atmp#1#2#3#4{\emph{#4}, \emph{ Adv. Theor. Math. Phys.} {\bf #1} (#3) #2}
\def\ptps#1#2#3#4{\emph{#4}, \emph{ Prog. Theor. Phys. Suppl.} {\bf #1} (#3) #2}
\def\lmp#1#2#3#4{\emph{#4}, \emph{ Lett. Math. Phys.} {\bf #1} (#3) #2}
\def\cpam#1#2#3#4{\emph{#4}, \emph{ Comm. Pure Appl. Math.}  {\bf #1} (#3) #2}
\def\adv#1#2#3#4{\emph{#4}, \emph{ Adv. Phys.}  {\bf #1} (#3) #2}
\def\zh#1#2#3#4{\emph{#4}, \emph{ Zh. Eksp. Teor. Fiz.}  {\bf #1} (#3) #2}

\def\jams#1#2#3#4{\emph{#4}, \emph{ J. Austral. Math. Soc. B} {\bf #1} (#3) #2}
\def\appa#1#2#3#4{\emph{#4}, \emph{ Acta Phys. Polonica A} {\bf #1}, (#3) #2}
\def\nat#1#2#3#4{\emph{#4}, \emph{Nature} {\bf #1}, (#3) #2}
\def\science#1#2#3#4{\emph{#4}, \emph{Science} {\bf #1}, (#3) #2}
\def\arcmp#1#2#3#4{\emph{#4}, \emph{Annual Rev. of Cond. Matter Physics} {\bf #1}, (#3) #2}
\def\jcap#1#2#3#4{\emph{#4}, \emph{JCAP} {\bf #1}, (#3) #2}
\def\conphy#1#2#3#4{\emph{#4}, \emph{Contemporary Physics} {\bf #1}, (#3) #2}
%
\def\hepph#1#2{{ hep-ph }{#1} (#2)}
\def\hepth#1#2{{ hep-th }{#1} (#2)}
\def\grqc#1#2{{ gr-qc }{#1} (#2)}
\def\ibid#1#2#3#4{\emph{#4}, {\it ibid.} {\bf #1} (#3) #2}
%
\bibitem{mal}
J.M.Maldacena, \atmp{2}{231}{1998}{The large-N limit of superconformal field theories and supergravity}.

\bibitem{wit98}
E.Witten, \atmp{2}{253}{1998}{Anti-de-Sitter space and holography}.
\bibitem{gub98}
S.S.Gubser, I.R.Klebanov and A.M.Polyakov, \plb{428}{105}{1998}{Gauge theory correlators from noncritical string theory}.

\bibitem{rev1} 
J.P.Gauntlett, J.Sonner and T.Wiseman
\prl{103}{151601}{2009}{Holographic Superconductivity in M Theory}.

\bibitem{sachdev2012} 
S.Sachdev, \arcmp{3}{2012}{9}{What can gauge-gravity duality teach us about condensed matter physics?}.

\bibitem{har08}
S.A.Hartnoll, C.P.Herzog and G.T.Horowitz, \prl{101}{031601}{2008}{Building a holographic superconductor}.


\bibitem{che10}
J.W.Chen, Y.J.Kao, D.Maity, W.Y.Wen and C.P.Yeh, \prd{81}{106008}{2010}{Towards a holographic model of D-wave superconductors}.
\bibitem{ben10}
F.Benini, C.P.Herzog, R.Rahman and A.Yarom, \jhep{11}{2010}{137}{Gauge gravity duality for d-wave superconductors: prospects and chalanges}.
\bibitem{rog14a}
M.Rogatko and K.I.Wysoki\'nski, \appa{126}{A9}{2014} {Remarks on the Hall conductivity in chiral superconductors: weak vs. strong coupling approach}.
\bibitem{zen10}
H.B.Zeng, Z.Y.Fan and H.S.Zong, \prd{82}{126008}{2010}{d-wave holographic superconductor vortex lattice and non-Abelian holographic superconductor droplet}.


\bibitem{gub08}
S.S.Gubser and S.S.Pufu, \jhep{11}{2008}{033}{The gravity dual of a p-wave superconductor}.
\bibitem{bas10}
P.Basu, J.He, A.Mukharjee and H.H.Shieh, \plb{689}{45}{2010}{Hard-gapped holographic superconductors}.
\bibitem{apr11}                  
F.Aprile, D.Rodriguez-Gomez and J.G.Russo, \jhep{01}{2011}{056}{p-wave holographic superconductors and five-dimensional gaiged supergravity}.
\bibitem{gan12}
S.Gangopadhyay and D.Roychowdhury, \jhep{08}{2012}{104}{Analytic study of properties of holographic p-wave superconductors}.
\bibitem{amm10}
M.Ammon, J.Erdmenger, V.Grass and P.Kerner, \plb{686}{192}{2010}{On holographic p-wave superfluids with back-reaction}.
\bibitem{liu15}
S.Liu, Y.Q.Wang, {\it Holographic model of hybrid and coexisting s-wave and p-wave 
Josephson junction}, \hepth{1504.0691}{2015}.



\bibitem{hor98}
G.T.Horowitz and R.C.Myers, \prd{59}{026005}{1998}{The AdS/CFT correspondence and a new positive energy conjecture for general relativity}.
\bibitem{nis10}
T.Nishioka, S.Ryu and T.Takayanagi, \jhep{03}{2010}{131}{Holographic superconductor/ insulator transition at zero temperature}.
\bibitem{wit98a}
E.Witten, \atmp{2}{505}{1998}{Anti-de Sitter space, thermal phase transition and confinement in gauge theories}.
\bibitem{hor10}
G.T.Horowitz and B.Way, \jhep{11}{2010}{011}{Complete phase diagrams for a holographic superconductor}.

\bibitem{bri11} 
Y.Brihaye and B.Hartmann, \prd{83}{126008}{2011}{Holographic superfluid/fluid/insulator phase transitions in 2+1 dimensions}.
\bibitem{cai11a}
R.G.Cai, X.He, H.F.Li and H.Q.Zhang, \prd{84}{046001}{2011}{Phase transitions in AdS soliton spacetime through marginally stable modes}.
\bibitem{cai11b}
R.G.Cai, L.Li, H.Q.Zhang and Y.L.Zhang, \prd{84}{126008}{2011}{Magnetic field effect on the phase transition in AdS soliton spacetime}.
\bibitem{cai11c}
R.G.Cai, H.F.Li, H.Q.Zhang, \prd{83}{126007}{2011}{Analytical studies on holographic insulator/superconductor phase transitions}.
\bibitem{akh11}
A.Akhavan and M.Alishahiha, \prd{83}{086003}{2011}{p-wave holographic insulator/superconductor phase transition}.

\bibitem{pan11}
Q.Pan, J.Jing and B.Wang, \jhep{11}{2011}{088}{Analytical investigation of the phase transition between holographic insulator and superconductor in Gauss-Bonnet gravity}.
\bibitem{cai13}
R.G.Cai, S.He, L.Li and L.F.Li, \jhep{12}{2013}{036}{A holographic study on vector condensate induced by a magnetic field}.
\bibitem{zha15}
L.Zhang, Q.Pan and J.Jing, \plb{743}{104}{2015}{Holographic p-wave superconductor models with Weyl corrections}.
\bibitem{cha15}
P.Chaturvedi and G.Sengupta, \jhep{04}{2015}{001}{p-wave holographic superconductors from Born-Infeld black holes}.

\bibitem{cai11back}
R.G.Cai, Z.Y.Nie and H.Q.Zhang, \prd{83}{066013}{2011}{Holographic phase transitions of p-wave superconductors in Gauss-Bonnet gravity with backreaction}.

\bibitem{zha13}
Z.Zhao, Q.Pan and J.Jing, \plb{719}{440}{2013}{ Holographic insulator/superconductor phase transition with Weyl corrections}.
\bibitem{jin12}
J.Jing, Q.Pan and S.Chen, \plb{716}{385}{2012}{Holographic superconductor/insulator transition with logarithmic electromagnetic field in Gauss–Bonnet gravity}.

\bibitem{alb09}
T.Albash and C.V.Johnson, \prd{80}{126009}{2009}{Vortex and droplet engineering in holographic superconductors}.
\bibitem{roy13}
D.Roychowdhury, \jhep{05}{2013}{162}{Holographic droplets in p-wave insulator/superconductor transition}.

\bibitem{amo14}
A.Amoretti, A.Braggio, N.Maggiore, N.Magnoli and D.Musso, \jhep{01}{2014}{054}{Coexistence of two vector order parameters: a holographic model for ferromagnetic superconductivity}.



\bibitem{integral}
P.Jean {\it et al.}, \aa{407}{L55}{2003}{Early SPI/INTEGRAL measurements of 511 keV line emission from the 4th quadrant of the Galaxy}.
\bibitem{atic}
J.Chang {\it et al.}, \nat{456}{362}{2008}{An excess of cosmic ray electrons at energies of 300-800 GeV}.
\bibitem{pamela}
O.Adriani {\it et al.} (PAMELA Collaboration), \nat{458}{607}{2009}{An anomalous positron abundance in cosmic rays with energies 1.5-100 Gev}.
\bibitem{massey15a}
D.Harvey, R.Massey, T.Kitching, A.Taylor and E.Tittley, \science{347}{1462}{2015}{The nongravitational interactions of dark matter in colliding galaxy clusters}.
\bibitem{massey15b}
R.Massey {\it et al.}, \mnras{449}{3393}{2015}{The behaviour of dark matter associated with four bright cluster galaxies in the 10 kpc core of Abell 3827}.



\bibitem{muon}
G.W.Bennett {\it et al.}, \prd{73}{072003}{2006}{Final report of the E821 muon anomalous magnetic moment measurement at BNL}.



\bibitem{afa09}
A.Afanasev, O.K. Baker, K.B.Beard, G.Biallas, J.Boyce, M.Minarni, R.Ramdon, M.Shinn, and P.Slocum, \plb{679}{317}{2009}{New experimental limit on photon hidden-sector paraphoton mixing}.
\bibitem{gni08}
S. N. Gninenko and J.Redondo, \plb{664}{180}{2008}{On search for EV hidden-sector in Supper-Kanionkande and CAST experiments}.
\bibitem{suz15}
J.Suzuki, T.Horie, Y.Inoue, and M.Minowa, \jcap{09}{042}{2015}{Experimental search for hidden photon CDM in the eV mass range with a dish antenna}.
\bibitem{mir09}
A.Mirizzi, J.Redondo, and G.Sigl, \jcap{03}{026}{2009}{Microwave background constraints on mixing of photons with hidden photons}.
\bibitem{red13}
J.Redondo and G.Raffelt, \jcap{08}{034}{2013}{Solar constraints on hidden photons re-visited}.






\bibitem{nak14}
{\L}.Nakonieczny and M.Rogatko, \prd{90}{106004}{2014}{Analytic study on backreacting holographic superconductors with dark matter sector}.
\bibitem{nak15}
{\L}.Nakonieczny, M.Rogatko and K.I. Wysoki\'nski, \prd{91}{046007}{2015}{Magnetic field in holographic superconductors with dark matter sector}.

\bibitem{nak15a} \L{}. Nakonieczny, M. Rogatko and K.I. Wysoki\'nski, \prd{92}{066008}{2015}
{\it Analytic investigation of holographic phase transitions influenced by
dark matter sector}.


\bibitem{vac91}
T.Vachaspati and A.Achucarro, \prd{44}{3067}{1991}{Semilocal cosmic strings}.
\bibitem{ach00}
A.Achucarro and T.Vachaspati, \prep{327}{347}{2000}{Semilocal and electroweak strings}.
\bibitem{har09}
B.Hartmann and F.Arbabzadah, \jhep{07}{2009}{068}{Cosmic strings interacting with dark strings}.
\bibitem{bri09}
Y.Brihaye and B.Hartmann, \prd{80}{123502}{2009}{Effect of dark strings on semilocal strings}.
\bibitem{dav12}
H.Davoudiasl, H.S.Lee and W.J.Marciano, \prd{85}{115019}{2012}{"Dark" Z implications for parity violation, rare meson decays, and Higgs physics}.
\bibitem{dav13}
H.Davoudiasl, H.S.Lee, I.Lewis and W.J.Marciano, \prd{88}{015022}{2013}{Higgs decays as a window into the dark sector}.


\bibitem{hel1}
A.Donos and J.P.Gauntlett, \jhep{12}{2011}{091}{Holographic helical superconductors}.
\bibitem{hel2}
A.Donos and J.P.Gauntlett, \prl{108}{211601}{2012}{Helical superconducting black holes}.
\bibitem{cai15}
R.G.Cai, L.Li, L.F.Li and R.Q.Yang, {\it Introduction to Holographic Superconductor Models},
\hepth{1502.00437}{2015}.

\bibitem{dju05}
D.Djukanovic, M.R.Schindler, J.Gegelia and S.Scherer, \prl{95}{012001}{2005}{Quantum electrodynamics for vector mesons}.




\bibitem{sio10}
G.Siopsis and J.Therrien, \jhep{05}{2010}{013}{Analytic calculation of properties of holographic superconductors}.

\bibitem{li11}
H.F.Li, R.G.Cai and H.Q.Zhang, \jhep{04}{2011}{028}{Analytical studies on holographic superconductors in Gauss-Bonnet gravity}.

\bibitem{gre09}
R.Gregory, S.Kanno and J.Soda, \jhep{10}{2009}{010}{Holographic superconductors with higher curvature corrections}.
\bibitem{cai10}
R.G.Cai, Z.Y.Nie and H.Q.Zhang, \prd{82}{066007}{2010}{Holographic p-wave superconductors from Gauss-Bonnet gravity}.
\bibitem{pan10}
Q.Pan, B.Wang, E.Papantonopoulos, J.Oliveira and A.Pavan, \prd{81}{106007}{2010}{Holographic superconductors with various condensates in Einstein-Gauss-Bonnet gravity}.

\bibitem{gre13} 
A.G. Green, \conphy{54}{33}{2013}{ An Introduction to Gauge Gravity Duality and Its Application in Condensed Matter}.

\bibitem{freese2013} K. Freese, M. Lisanti, Ch. Savage, {\it Colloquium: Annual modulation of dark matter},
Rev. Mod. Physics {\bf 85} (2013) 1561. 

\end{thebibliography}
\end{document}